# Interference Networks with General Message Sets: A Random Coding Scheme

Reza Khosravi-Farsani, Farokh Marvasti

Adnaced Communications research Institue (ACRI)
Department of Electrical Engineering, Sharif University of Technology, Tehran, Iran
Email: reza_khosravi@alum.sharif.ir, marvasti@sharif.ir

*Abstract*—In this paper, the Interference Network with General Message Sets (IN-GMS) is introduced in which several transmitters send messages to several receivers: Each subset of transmitters transmit an individual message to each subset of receivers. For such a general scenario, an achievability scheme is presented using the random coding. This scheme is systematically built based on the capacity achieving scheme for the Multiple Access Channel (MAC) with common message as well as the best known achievability scheme for the Broadcast Channel (BC) with common message. A graphical illustration of the random codebook construction procedure is also provided, by using which the achievability scheme is easily understood. Some benefits of the proposed achievability scheme are described. It is also shown that the resulting rate region is optimal for a class of orthogonal INs-GMS, which yields the capacity region. Finally, it is demonstrated that how this general achievability scheme can be used to derive capacity inner bounds for interference networks with different distribution of messages; in most cases, the proposed achievability scheme leads to the best known capacity inner bound for the underlying channel. Capacity inner bounds can also be derived for new communication scenarios.

## I. INTRODUCTION

The interference networks are of the most important multiuser scenarios due to the wide range of practical communications systems for which these models are fitted. Up to know these networks have been extensively studied, however, our knowledge regarding the behavior of information flow in them is still limited. For instance, a computable characterization of the capacity region for the two-user Classical Interference Channel (CIC) is unknown [1], unlike its simple configuration. The best achievability scheme for this channel is due to Han-Kobayashi (HK) [2] proposed in 1981. The multiuser interference networks recently have been widely investigated in the literature. Nevertheless, they are far less understood [3].

In this paper, we introduce the Interference Networks with General Message Sets (IN-GMS), a network scenario where several transmitters send messages to several receivers: Each subset of transmitters transmit an individual message to each subset of receivers. In fact, this scenario unifies all interference channel models with diverse distribution of messages. For example, the two-transmitter/two-receiver IN-GMS contains the CIC, the Multiple Access Channel (MAC) with common message [4], the Broadcast Channel (BC) with common message, the cognitive radio channel [5], the X-channel [6], the Z-channel [7], the cognitive interference channel with degraded message sets [8] and etc, as special cases. In this paper, we present a random coding scheme for such a general scenario. Having at hand an achievable rate region for this channel in the general case sheds light on information flow, not only for the system itself but also for its sub-channels. Specifically, we demonstrate that all previously derived achievable rate regions for different interference networks can be deduced from our general scheme [9].

To building achievability schemes with satisfactory performance for such large networks, it is required:

1. To recognize the main building blocks involved in the network.
2. To know the best encoding/decoding strategy for each building block.
3. To combine *systematically* the best achievability schemes of the building blocks.

In this paper, regarding the first step, we justify that the MAC with common messages and the BC with common messages are two main building blocks of the IN-GMS, which should be focused on to derive a high performance achievability scheme for this network. We then discuss in details the best encoding/decoding strategy for these two models. Precisely speaking, for the MAC with common messages it was shown [4] that superposition coding achieves the capacity. Regarding the BC, the capacity region is still unknown; the best achievability scheme for the two-user BC is due to Marton [10]. In this paper, we provide a graphical illustration for the superposition structures among the generated codewords in a random coding scheme, by which the encoding procedure is easily understood. Based on this graphical representation, we argue that the superposition structures among the generated codewords in the Marton's coding for the two-user BC with common message is exactly the same as that one in the MAC with common message. We examine some other coding strategies for the two-user BC and mention that the resulting achievable rate region by them is equivalent to Marton's one or include in it as its subsets. Using these general insights, we propose a random coding scheme for the multi-receiver BC with common messages (for each subset of the receivers there exist a common message), in which the superposition structures among the generated codewords are exactly similar to the multi-transmitter MAC with common messages.

As the last step, we combine systematically these two encoding strategies, i.e., the capacity achieving scheme for the MAC with common messages and the proposed coding for the BC with common messages, to building an achievability scheme for the IN-GMS. As one of the useful properties our achievability scheme is that the superposition structures among the RVs is such that each receiver decodes only its respective messages (using a jointly typical decoder) and it is not required to decode non-intended messages at some receivers. We also demonstrate that our achievable rate region is optimal for a class of orthogonal IN-MAC. Then, we describe that how our general achievability scheme can be used to derive capacity inner bounds for interference networks with diverse distribution of messages.

It should be mentioned that due to simplicity of exposition, in this conference version of our paper, we only discuss the achievability scheme for the two-transmitter/two-receiver case; nevertheless, our systematic approach is such that all the rules in derivation of the coding scheme directly extend to the case with arbitrary number of transmitters and receivers, as will be reported in [9]. Moreover, to analyze the error probability of the proposed coding, we exploit a covering lemma proved in [3, p. 15-40]. Using a novel application of this lemma the necessary conditions for vanishing the error probability in the encoding steps are readily derived, which makes the analysis significantly concise. Also, the analysis of the decoding steps is performed by constructing a table of decoding errors, in a clear framework with a few computations.

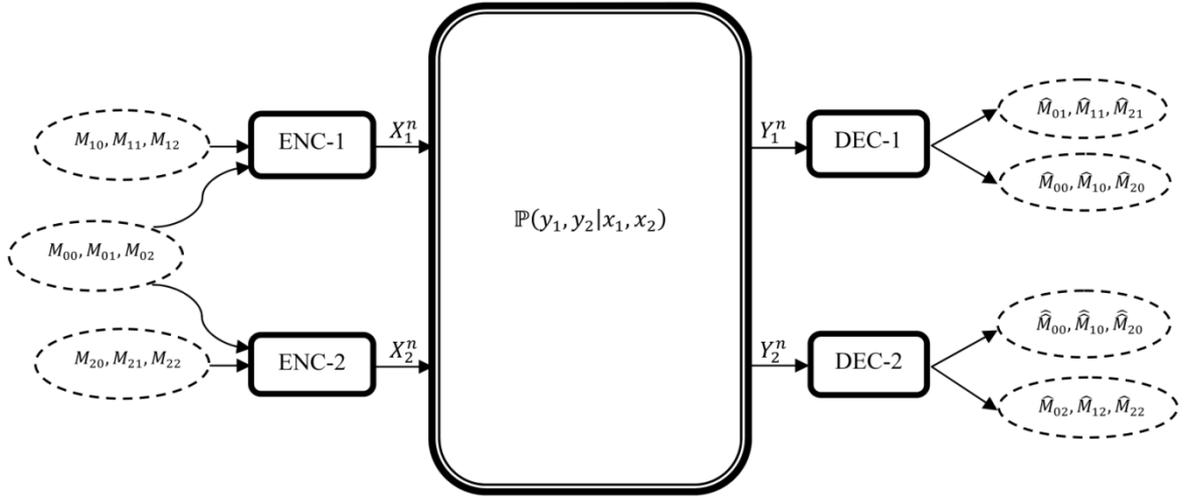

Figure 1. The two-transmitter/two-receiver Interference Network with General Message Sets (IN-GMS).

In the rest of the paper, we briefly state the preliminaries and channel model definitions in Section II. The main results are given in Section III. Due to space limitations, some steps in the analysis of the coding scheme are omitted here, but they can be found in [11]. The generalization of the coding scheme for networks with arbitrary number of transmitters and receivers is given in [9].

## II. PRELIMINARIES AND DEFINITIONS

Throughout the paper the following notations are used: Random Variables (RV) are denoted by upper case letters (e.g. $X$) and lower case letters are used to show their realization (e.g. $x$). The range set of a RV $X$ is represented by $\mathcal{X}$. The Probability Distribution Function (PDF) of a RV $X$ is denoted by $P_X(x)$, and the conditional PDF of $X$ given $Y$ is denoted by $P_{X|Y}(x|y)$; also, in representing PDFs, the arguments are sometimes omitted for brevity. The probability of the event $A$ is expressed by $Pr(A)$. The set of nonnegative real numbers and positive integers are denoted by $\mathbb{R}_+$, and $\mathbb{Z}_+$, respectively. The notation $[1:K]$ where $K$ is a positive integer, represents the set $\{1,\dots,K\}$. The set of all jointly $\epsilon$-letter typical $n$-sequences $(x^n, y^n)$ with respect to the PDF $P_{XY}(x, y)$ is denoted by $\mathcal{T}_\epsilon^n(P_{XY})$, (To see the definition of such sequences and their properties refer to [12]). Also, given the sequence $y^n$, the set of all $n$-sequences $x^n$ which are jointly typical with $y^n$ with respect to the PDF $P_{XY}(x, y)$, is denoted by $\mathcal{T}_\epsilon^n(P_{XY}|y^n)$. Finally, $p_{min}(P_X)$ denotes the minimum positive value of $P_X$.

**Interference Networks with General Message Sets:** Here, we briefly discuss the communications scenario of the IN-GMS in the two-transmitter/two-receiver case. The detailed definitions are given in [9] wherein the general network from the viewpoint of the number of transmitters and receivers is considered.

Consider a two-transmitter/two-receiver interference network wherein the transmitters intend to send nine messages over the channel; there exist three sets of triple messages where one message set is transmitted over the channel by both transmitters cooperatively, and the two other message sets are transmitted separately, one set by each transmitter. In each message set there exist three messages: two private messages, one for each receiver, and also a common message for both receivers. Therefore, each receiver is required to decode six messages three of which are common between both receivers. This channel indeed includes all possible schemes of transmitting messages over a two-user interference network. Hence, we refer to as *Interference Network with General Message Sets* (IN-GMS). Figure 1 illustrates the channel model.

This network is determined by the conditional PDF $\mathbb{P}(y_1, y_2 | x_1, x_2)$ which describes the relation between inputs and outputs of the network. The network is assumed to be memoryless. For a length-$n$ block code, $n \in \mathbb{Z}_+$, the $i^{th}$ transmitter encodes its respective messages using the codewords $X_i^n$ and the $j^{th}$ receiver decodes its intended messages by the received sequence $Y_j^n$, $i, j = 1,2$. The explicit definitions of the encoding and decoding procedures and the capacity region for the IN-GMS can be found in [9]. As usual, every subset of the capacity region of the network is called an achievable rate region.

In the next section, we present an achievable rate region for this network using the random coding.

## III. MAIN RESULTS

In this section, we aim at establishing an achievability scheme for the IN-GMS depicted in Fig. 1. Due to presence of several messages which are required to transmit over the channel, one can consider numerous achievability schemes for this network. But the question is that what is the best transmission strategy?

To respond to this question, first, we discuss the main building blocks involved in the network as well as the best encoding/decoding strategy for each one. To recognize the main building blocks of the IN-GMS, we look at the encoding and the decoding sides of the network. Let us examine Fig. 1. From the viewpoint of the encoding side, we have a multiple access problem with common message. On the other hand, from the viewpoint of the decoding side, we have a broadcasting problem, (both common and private messages). Therefore, it is required to investigate the MAC with common message and also the BC with common message, in details. Consider the MAC with a common message, as shown in Fig. 2.

The capacity region of this channel was determined in [4] which is given as:

$$\bigcup_{P_W P_{X_1|W} P_{X_2|W}} \begin{cases} (R_1, R_2) \in \mathbb{R}_+^2 : \\ R_1 \leq I(X_1; Y | X_2, W) \\ R_2 \leq I(X_2; Y | X_1, W) \\ R_1 + R_2 \leq I(X_1, X_2; Y | W) \\ R_0 + R_1 + R_2 \leq I(X_1, X_2; Y) \end{cases}$$

(1)

For this channel, it was shown that the superposition coding achieves the capacity. As a brief discussion regarding this coding scheme, we mention that the common message $M_0$ is encoded by a codeword constructed by the RV $W$ based on $P_W$. Then, for each of the private messages a codeword is generated superimposing on the common message codeword $W$: The private message $M_i$ is encoded using a codeword constructed by $X_i$ based on $P_{X_i|W}$, $i = 1,2$. The $i^{th}$ transmitter, $i = 1,2$, then sends $X_i(M_i, M_0)$ over the channel. The decoder decodes the messages using a jointly typical decoder. Figure 3 graphically illustrates the encoding scheme.

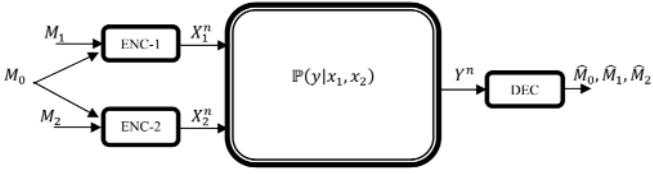

Figure 2. The two-user MAC with a common message.

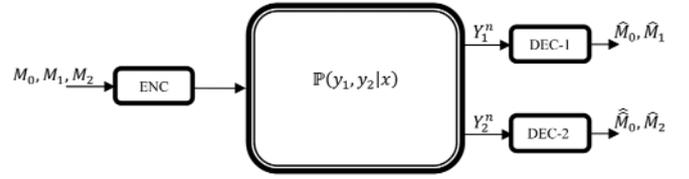

Figure 4. The two-user BC with common message.

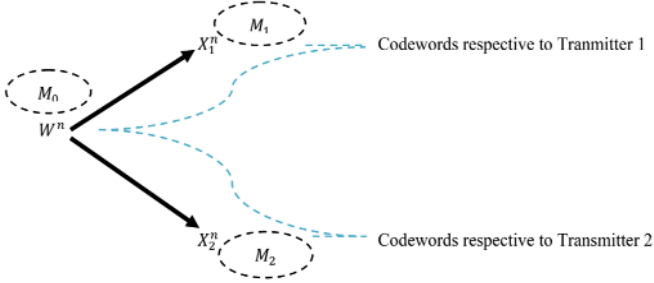

Figure 3. The graphical illustration of the generated codewords for the MAC with a common message. Every two codewords connected by an arrow build a superposition structure: The codeword at the beginning of the arrow is the cloud center and that one at the end of the arrow is the satellite. The ellipse beside each codeword shows what contains that codeword, in addition to those ones in its cloud centers.

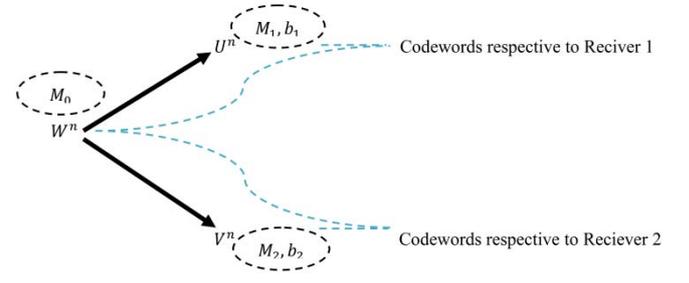

Figure 5. The graphical illustration of the generated codewords for the BC with a common message in the Marton's scheme. This figure depicts the superposition structures among the generated codewords. The parameters $b_1, b_2$ indicate the bin indices.

In this illustration, we use a directed graph to represent the superposition structures among the generated codewords: Every two codewords connected by an arrow (directed age) build a superposition structure where the codeword at the beginning of the arrow is the cloud center and that one at the end of the arrow is the satellite. The ellipse beside each codeword shows what contains that codeword, in addition to those ones in its cloud centers. This graphical representation is very useful to understand an achievability scheme, especially for large networks.

Then, consider the two-user BC with common information, as shown in Fig. 4. The capacity region of the BC is still an unsolved problem in network information theory. To date, the best capacity inner bound for this channel is due to Marton [10], (see also [13]) which is given by:

$$\bigcup_{P_{WUVX}} \begin{cases} (R_0, R_1, R_2) \in \mathbb{R}_+^3 : \\ R_0 + R_1 \leq I(W, U; Y_1) \\ R_0 + R_2 \leq I(W, V; Y_2) \\ R_0 + R_1 + R_2 \leq I(W, U; Y_1) + I(V; Y_2|W) \\ \qquad\qquad\qquad - I(U; V|W) \\ R_0 + R_1 + R_2 \leq I(U; Y_1|W) + I(W, V; Y_2) \\ \qquad\qquad\qquad - I(U; V|W) \\ 2R_0 + R_1 + R_2 \leq I(W, U; Y_1) + I(W, V; Y_2) \\ \qquad\qquad\qquad - I(U; V|W) \end{cases}$$
(2)

Here, we briefly discuss the Marton's coding scheme. Roughly speaking, the common message $M_0$ is encoded by a codeword constructed of $W$ based on $P_W$. For each of the private messages, a bin of codewords is randomly generated superimposing on the common message codeword $W$: The bin respective to $M_1$ is constructed by $U$ based on $P_{U|W}$ and that one for $M_2$ is constructed by $V$ based on $P_{V|W}$. These bins are explored against each other to find a jointly typical pair of codewords. Using the mutual covering lemma [3], the sizes of the bins are selected sufficiently large such that the existence of such typical pair of codewords is guaranteed. Superimposing on the designated codewords $W, U, V$, the encoder then generates its codewords constructed by $X$ based on $P_{X|WUV}$, and sends it over the channel. Each receiver decodes its respective codewords (the first one decodes $W, U$ and the second one decodes $W, V$) using a jointly typical decoder. The resulting achievable rate region is further enlarged and reaches to (2) by the fact that if the rate triple $(R_0, R_1, R_2) \in \mathbb{R}_+^3$ is achievable for the BC, then $(R_0 - \pi_1 - \pi_2, R_1 + \pi_1, R_2 + \pi_2) \in \mathbb{R}_+^3$ is also achievable. The graphical representation of the Marton's coding has been shown in Fig. 5.

The superposition structures among the generated codewords in the Marton's coding scheme for the two-user BC with common message are exactly the same as the MAC with common message, as shown in Fig. 3. The only difference in the encoding scheme is that, unlike the MAC, for the BC since all the messages are available at one transmitter, it is possible to apply the binning technique. Using the binning scheme we can construct the transmitted codewords jointly typical with the PDF $P_{WUVX}$, which yields a larger achievable rate region than the case where the messages are encoded only using the superposition coding according to the PDF $P_W P_{U|W} P_{V|W} P_{X|WUV}$.

It should also be mentioned that one can consider some new coding schemes for the two-user BC other than the Marton's one. For example, it is possible to encode all the messages (both common and private messages) only using the binning technique, i.e., without superposition coding. In this scheme, roughly speaking, respective to each message a bin of codewords is generated (the bins are generated independently) and then these three bins are explored against each other to find a jointly typical triple. Using the multivariate covering lemma [3] the sizes of the bins are selected such large to guarantee that there exists such triple of codewords. The transmitter then generates its codeword superimposing on this jointly typical triple and sends it over the channel. Each receiver decodes its respective messages using a jointly typical decoder. Other coding strategies are also available. We have examined these coding schemes [9] and found that all the resulting achievable rate regions are equivalent to the Marton's one or include in it as its subset. Therefore, we can conclude that to broadcasting both common and private messages, it is more beneficial to encode the private messages superimposing on the common messages.

*Using this general insight, in [9] we propose an achievability scheme for transmission of the general message sets over the multi-receiver BC such that the superposition structure among the generated codeword is exactly similar to the multi-transmitter MAC with common messages [4]. To derive this superposition structure it is sufficient to look at the receivers of the BC from the viewpoint of the respective messages, as the transmitters of a MAC. The details can be found in [9].*

The Marton's achievable rate region (2) for the two-user BC is optimal in all special cases for which the capacity region is known; specifically, the degraded BCs, the more-capable BCs, the semi-deterministic BCs [3]. It is also optimal for the Gaussian multiple-input multiple-output BCs [3].

Now, let us turn to the IN-GMS depicted in Fig. 1. In the following, we first derive an achievable rate region for this network and then we show that some known results, specifically, the HK rate region [2] for the two-user CIC can be derived from our coding scheme as special cases.

Consider the IN-GMS as depicted in Fig. 1. In this model, three sets of messages, i.e., $\{M_{10}, M_{11}, M_{12}\}, \{M_{00}, M_{01}, M_{02}\}, \{M_{20}, M_{21}, M_{22}\}$, are sent over the channel where from the view point of each set we have a broadcasting scenario: One private message for each receiver and a common message for both. As mentioned before, the main building blocks of the network are the two-user BC with common message and also the two-user MAC with common message. Therefore, to derive a satisfactory achievability scheme for this network, it is required to combine systematically the best coding schemes for these main building blocks. Note that by considering transmission of only one of the message sets $\{M_{10}, M_{11}, M_{12}\}, \{M_{00}, M_{01}, M_{02}\}, \{M_{20}, M_{21}, M_{22}\}$, the IN-GMS reduces to the two-user BC; therefore, we build our achievability scheme such that when it is specialized for these sub-channels, the Marton's inner bound (2) for the two-user BC results.

Note that here we describe our coding scheme in details only for the two-transmitter/two-receiver IN-GMS; nevertheless, due to our systematic approach, all the rules applied here to establish the achievability scheme directly extend to the case of arbitrary number of transmitters and receivers, as will be reported in [9]. Also, it is worth noting that, however, our achievability scheme may seem complex at the first glance, but indeed this is not the case. Due to symmetry in the encoding and decoding steps, the analysis of the proposed random coding is very simple. In addition, in the encoding steps we exploit a multivariate covering lemma proved in [3, p. 15-40] to obtain an admissible source region for the two-user BC. Using a novel application of this lemma, the necessary conditions for vanishing error probability in encoding steps are readily derived, which makes the analysis significantly concise; see [11] for details. In the following theorem we state our main result.

**Theorem 1)** Define the rate region $\mathfrak{R}_i^{IN-GMS}$ as given in the next page. The set $\mathfrak{R}_i^{IN-GMS}$ is an achievable rate region for the IN-GMS depicted in Fig. 1.

*Remarks:*

1. The rate region $\mathfrak{R}_i^{IN-GMS}$ is convex.

2. The rate region $\mathfrak{R}_i^{IN-GMS}$ can be further enlarged by considering the fact that if $(R_{00}, R_{01}, R_{02}, R_{10}, R_{11}, R_{12}, R_{20}, R_{21}, R_{22}) \in \mathbb{R}_+^9$ is achievable, then the following 9-tuples are also achievable:

$(R_{00} - \pi_1 - \pi_2, R_{01} + \pi_1, R_{02} + \pi_2, R_{10}, R_{11}, R_{12}, R_{20}, R_{21}, R_{22})$

$(R_{00}, R_{01}, R_{02}, R_{10} - \pi_1 - \pi_2, R_{11} + \pi_1, R_{12} + \pi_2, R_{20}, R_{21}, R_{22})$

$(R_{00}, R_{01}, R_{02}, R_{10}, R_{11}, R_{12}, R_{20} - \pi_1 - \pi_2, R_{21} + \pi_1, R_{22} + \pi_2)$

where $(\pi_1, \pi_2, \pi_2) \in \mathbb{R}_+^3$. This fact is adapted from the same observation for the BC, as discussed before.

3. One of the useful properties our achievability scheme is that the superposition structures among the RVs is such that each receiver decodes only its respective messages (using a jointly typical decoder) and it is not required to decode non-intended messages at some receivers. This is important, since usually decoding non-intended messages at one receiver causes rate loss.

For the special cases of the two-user MAC with common message and the BC with common message our achievable rat region (after applying the technique mentioned in Remark 2) reduces to the (1) and (2), respectively.

*Proof of Theorem 1)*

We derive the achievability of $\mathfrak{R}_i^{IN-GMS}$ given by (7) using a random coding argument. To encode each of the messages $\{M_{10}, M_{11}, M_{12}\} \cup \{M_{00}, M_{01}, M_{02}\} \cup \{M_{20}, M_{21}, M_{22}\}$, we use an auxiliary RV. Inspired by Marton's region characterization given by (2), we encode the messages $M_{i0}, M_{i1}, M_{i2}, i = 0,1,2$, by $W_i, U_i, V_i$, respectively.

*Definition:* Suppose $m \in \mathbb{Z}_+$. Let $\Lambda_m: \mathbb{Z}_+^m \to \mathbb{Z}_+$ be a bijection. The *order* relation $<_{\Lambda_m}$ induced by $\Lambda_m(.)$ on the set $\mathbb{Z}_+^m$, is defined as follows. For every $(a_1, \ldots, a_m)$ and $(b_1, \ldots, b_m)$ in $\mathbb{Z}_+^m$ where $(a_1, \ldots, a_m) \neq (b_1, \ldots, b_m)$, we have:

$$(a_1, \ldots, a_m) <_{\Lambda_m} (b_1, \ldots, b_m) \Leftrightarrow$$
$$\Lambda_m(a_1, \ldots, a_m) < \Lambda_m(b_1, \ldots, b_m) \quad (3)$$

Also, the "min" operator with respect to $<_{\Lambda_m}$, denoted by $\min \Lambda_m$, is defined as follows. Let $S$ be a nonempty subset of $\mathbb{Z}_+^m$. We have:

$$\min \Lambda_m S \triangleq \Lambda_m^{-1}(\min \{\Lambda_m(s): s \in S\}) \quad (4)$$

where $\Lambda_m^{-1}(.)$ denotes the inverse function. The "max" operator could be defined, similarly.

Let $(R_{00}, R_{01}, R_{02}, R_{10}, R_{11}, R_{12}, R_{20}, R_{21}, R_{22}) \in \mathbb{R}_+^9$, and the message $M_{ij}, i, j = 0,1,2$, be a RV uniformly distributed over the set $[1: 2^{nR_{ij}}]$. Also, let $\Lambda_2: \mathbb{Z}_+^2 \to \mathbb{Z}_+$ and $\Lambda_3: \mathbb{Z}_+^3 \to \mathbb{Z}_+$ be two arbitrary bejections with the "min" operators $\min \Lambda_2$ and $\min \Lambda_3$, respectively, as defined by (4). As a convention, denote $\min \Lambda_2(\emptyset) \triangleq (1,1)$ and $\min \Lambda_3(\emptyset) \triangleq (1,1,1)$.

*Encoding steps:* The encoding is performed in three steps:

*Step 1:* At the first step the messages $\{M_{00}, M_{01}, M_{02}\}$ which are sent by both transmitters cooperatively, are encoded. These messages are encoded exactly similar to Marton's coding scheme: Fix the PDFs $P_{W_0}, P_{U_0|W_0}, P_{V_0|W_0}$. Let $(B_{01}, B_{02}) \in \mathbb{R}_+^2$ be a nonnegative pair of real numbers. These serve as the sizes of the bins.

1. Generate at random $2^{nR_{00}}$ independent codewords $W_0^n$ according to $Pr(w_0^n) = \prod_{t=1}^n P_{W_0}(w_{0,t})$. Label these codewords $W_0^n(m_{00})$, where $m_{00} \in [1: 2^{nR_{00}}]$.

2. For each $W_0^n(m_{00})$, where $m_{00} \in [1: 2^{nR_{00}}]$, randomly generate $2^{n(R_{01}+B_{01})}$ independent codewords $U_0^n$ according to $Pr(u_0^n) = \prod_{t=1}^n P_{U_0|W_0}(u_{0,t}|w_{0,t}(m_{00}))$. Label these codewords $U_0^n(m_{00}, m_{01}, b_{01})$, where $m_{01} \in [1: 2^{nR_{01}}]$ and $b_{01} \in [1: 2^{nB_{01}}]$.

3. For each $W_0^n(m_{00})$, where $m_{00} \in [1: 2^{nR_{00}}]$, randomly generate $2^{n(R_{02}+B_{02})}$ independent codewords $V_0^n$ according to $Pr(v_0^n) = \prod_{t=1}^n P_{V_0|W_0}(v_{0,t}|w_{0,t}(m_{00}))$. Label these codewords $V_0^n(m_{00}, m_{02}, b_{02})$, where $m_{02} \in [1: 2^{nR_{02}}]$ and $b_{02} \in [1: 2^{nB_{02}}]$.

Given $m_{00}, m_{01}, m_{02}$, define the pair $(b_{01}^{\mathcal{T}}, b_{02}^{\mathcal{T}})$ as follows:

$$(b_{01}^{\mathcal{T}}, b_{02}^{\mathcal{T}}) \triangleq \min \Lambda_2 \left\{ \begin{array}{l} (b_{01}, b_{02}), b_{0i} \in [1: 2^{nB_{0i}}], i = 1,2: \\ \begin{pmatrix} W_0^n(m_{00}), \\ U_0^n(m_{00}, m_{01}, b_{01}), \\ V_0^n(m_{00}, m_{02}, b_{02}) \end{pmatrix} \in \mathcal{T}_\epsilon^n \end{array} \right\} \quad (5)$$

In other words, $(b_{01}^{\mathcal{T}}, b_{02}^{\mathcal{T}})$ is the minimum pair $(b_{01}, b_{02})$ (with respect to $\Lambda_2$) such that the codewords $W_0^n, U_0^n, V_0^n$ are jointly typical. If there is no such codewords, then $(b_{01}^{\mathcal{T}}, b_{02}^{\mathcal{T}}) \triangleq (1,1)$.

In the first step, the designated codewords for transmission are:

$$\left(W_0^n(m_{00}), U_0^n(m_{00}, m_{01}, b_{01}^{\mathcal{T}}), V_0^n(m_{00}, m_{02}, b_{02}^{\mathcal{T}})\right) \quad (6)$$

Using the mutual covering lemma [3], we can select the sizes of the bins $B_{01}, B_{02}$ sufficiently large to guarantee that the codewords (6) are jointly typical with respect to $P_{W_0 U_0 V_0}$.

In the next two steps, the two message sets $\{M_{10}, M_{11}, M_{12}\}$ and $\{M_{20}, M_{21}, M_{22}\}$ which are sent by transmitter 1 and 2, respectively, are encoded. The codewords generated in Step 1 are now served as cloud centers for the new codewords (which are generated in Steps 2 and 3) in such a fashion as depicted in Fig. 6.

$$\mathfrak{R}_i^{IN-GMS} \triangleq \bigcup_{\mathcal{P}_i^{IN-GMS}} \left\{ \begin{aligned}
&(R_{00}, R_{01}, R_{02}, R_{10}, R_{11}, R_{12}, R_{20}, R_{21}, R_{22}) \in \mathbb{R}_+^9: \\
&\exists\, (B_{01}, B_{02}, B_{10}, B_{11}, B_{12}, B_{20}, B_{21}, B_{22}) \in \mathbb{R}_+^8, \\
&\quad B_{01} + B_{02} \geq I(U_0; V_0|W_0) \\
&\quad\quad B_{i0} \geq I(U_0, V_0; W_i|W_0), i=1,2 \\
&\quad\quad B_{i0} + B_{i1} \geq I(U_0, V_0; W_i|W_0) + I(V_0; U_i|W_0, U_0, W_i), i=1,2 \\
&\quad\quad B_{i0} + B_{i2} \geq I(U_0, V_0; W_i|W_0) + I(U_0; V_i|W_0, V_0, W_i), i=1,2 \\
&\quad B_{i0} + B_{i1} + B_{i2} \geq I(U_0, V_0; W_i|W_0) + I(V_0; U_i|W_0, U_0, W_i) + I(U_0, U_i; V_i|W_0, V_0, W_i), i=1,2 \\
&\quad R_{ij}^b = R_{ij} + B_{ij}, \quad i,j \in \{0,1,2\}, (i,j) \neq (0,0), \\
&\quad R_{11}^b < I_{E_1^d \to Y_1}, \quad R_{12}^b < I_{E_1^d \to Y_2} \\
&\quad R_{21}^b < I_{E_2^d \to Y_1}, \quad R_{22}^b < I_{E_2^d \to Y_2} \\
&\quad R_{11}^b + R_{21}^b < I_{E_3^d \to Y_1}, \quad R_{12}^b + R_{22}^b < I_{E_3^d \to Y_2} \\
&\quad R_{10}^b + R_{11}^b < I_{E_4^d \to Y_1}, \quad R_{10}^b + R_{12}^b < I_{E_4^d \to Y_2} \\
&\quad R_{20}^b + R_{21}^b < I_{E_5^d \to Y_1}, \quad R_{20}^b + R_{22}^b < I_{E_5^d \to Y_2} \\
&\quad R_{10}^b + R_{11}^b + R_{21}^b < I_{E_6^d \to Y_1}, \quad R_{10}^b + R_{12}^b + R_{22}^b < I_{E_6^d \to Y_2} \\
&\quad R_{11}^b + R_{20}^b + R_{21}^b < I_{E_7^d \to Y_1}, \quad R_{12}^b + R_{20}^b + R_{22}^b < I_{E_7^d \to Y_2} \\
&\quad R_{01}^b + R_{11}^b + R_{21}^b < I_{E_8^d \to Y_1}, \quad R_{02}^b + R_{12}^b + R_{22}^b < I_{E_8^d \to Y_2} \\
&\quad R_{10}^b + R_{11}^b + R_{20}^b + R_{21}^b < I_{E_9^d \to Y_1}, \quad R_{10}^b + R_{12}^b + R_{20}^b + R_{22}^b < I_{E_9^d \to Y_2} \\
&\quad R_{01}^b + R_{10}^b + R_{11}^b + R_{21}^b < I_{E_{10}^d \to Y_1}, \quad R_{02}^b + R_{10}^b + R_{12}^b + R_{22}^b < I_{E_{10}^d \to Y_2} \\
&\quad R_{01}^b + R_{11}^b + R_{20}^b + R_{21}^b < I_{E_{11}^d \to Y_1}, \quad R_{02}^b + R_{12}^b + R_{20}^b + R_{22}^b < I_{E_{11}^d \to Y_2} \\
&\quad R_{01}^b + R_{10}^b + R_{11}^b + R_{20}^b + R_{21}^b < I_{E_{12}^d \to Y_1}, \quad R_{02}^b + R_{10}^b + R_{12}^b + R_{20}^b + R_{22}^b < I_{E_{12}^d \to Y_2} \\
&\quad R_{00} + R_{01}^b + R_{10}^b + R_{11}^b + R_{20}^b + R_{21}^b < I_{E_{13}^d \to Y_1}, \quad R_{00} + R_{02}^b + R_{10}^b + R_{12}^b + R_{20}^b + R_{22}^b < I_{E_{13}^d \to Y_2}
\end{aligned} \right\}$$

(7)

where,

$$\begin{aligned}
&\theta_{Y_1}^1 = I(U_0; W_1|W_0), \quad \theta_{Y_2}^2 = I(U_0; W_2|W_0), \quad \theta_{Y_1}^3 = I(W_1; W_2|W_0, U_0), \quad \theta_{Y_1}^4 = I(U_0; W_1, W_2|W_0) \\
&\theta_{Y_1}^5 = I(U_1; W_2|W_0, U_0, W_1) \quad \theta_{Y_1}^6 = I(U_2; W_1|W_0, U_0, W_2), \quad \theta_{Y_1}^7 = I(U_1; U_2|W_0, U_0, W_1, W_2)
\end{aligned}$$

(8)

$$\begin{aligned}
I_{E_1^d \to Y_1} &= I(U_1; Y_1|W_0, U_0, W_1, W_2, U_2) + \theta_{Y_1}^5 + \theta_{Y_1}^7 \\
I_{E_2^d \to Y_1} &= I(U_2; Y_1|W_0, U_0, W_1, W_2, U_1) + \theta_{Y_1}^6 + \theta_{Y_1}^7 \\
I_{E_3^d \to Y_1} &= I(U_1, U_2; Y_1|W_0, U_0, W_1, W_2) + \theta_{Y_1}^5 + \theta_{Y_1}^6 + \theta_{Y_1}^7 \\
I_{E_4^d \to Y_1} &= I(W_1, U_1; Y_1|W_0, U_0, W_2, U_2) + \theta_{Y_1}^1 + \theta_{Y_1}^3 + \theta_{Y_1}^5 + \theta_{Y_1}^6 + \theta_{Y_1}^7 \\
I_{E_5^d \to Y_1} &= I(W_2, U_2; Y_1|W_0, U_0, W_1, U_1) + \theta_{Y_1}^2 + \theta_{Y_1}^3 + \theta_{Y_1}^5 + \theta_{Y_1}^6 + \theta_{Y_1}^7 \\
I_{E_6^d \to Y_1} &= I(W_1, U_1, U_2; Y_1|W_0, U_0, W_2) + \theta_{Y_1}^1 + \theta_{Y_1}^3 + \theta_{Y_1}^5 + \theta_{Y_1}^6 + \theta_{Y_1}^7 \\
I_{E_7^d \to Y_1} &= I(U_1, W_2, U_2; Y_1|W_0, U_0, W_1) + \theta_{Y_1}^2 + \theta_{Y_1}^3 + \theta_{Y_1}^5 + \theta_{Y_1}^6 + \theta_{Y_1}^7 \\
I_{E_8^d \to Y_1} &= I(U_0, U_1, U_2; Y_1|W_0, W_1, W_2) + \theta_{Y_1}^4 + \theta_{Y_1}^5 + \theta_{Y_1}^6 + \theta_{Y_1}^7 \\
I_{E_9^d \to Y_1} &= I(W_1, U_1, W_2, U_2; Y_1|W_0, U_0) + \theta_{Y_1}^1 + \theta_{Y_1}^2 + \theta_{Y_1}^3 + \theta_{Y_1}^5 + \theta_{Y_1}^6 + \theta_{Y_1}^7 \\
I_{E_{10}^d \to Y_1} &= I(U_0, W_1, U_1, U_2; Y_1|W_0, W_2) + \theta_{Y_1}^1 + \theta_{Y_1}^2 + \theta_{Y_1}^3 + \theta_{Y_1}^5 + \theta_{Y_1}^6 + \theta_{Y_1}^7 \\
I_{E_{11}^d \to Y_1} &= I(U_0, U_1, W_2, U_2; Y_1|W_0, W_1) + \theta_{Y_1}^1 + \theta_{Y_1}^2 + \theta_{Y_1}^3 + \theta_{Y_1}^5 + \theta_{Y_1}^6 + \theta_{Y_1}^7 \\
I_{E_{12}^d \to Y_1} &= I(U_0, W_1, U_1, W_2, U_2; Y_1|W_0) + \theta_{Y_1}^1 + \theta_{Y_1}^2 + \theta_{Y_1}^3 + \theta_{Y_1}^5 + \theta_{Y_1}^6 + \theta_{Y_1}^7 \\
I_{E_{13}^d \to Y_1} &= I(W_0, U_0, W_1, U_1, W_2, U_2; Y_1) + \theta_{Y_1}^1 + \theta_{Y_1}^2 + \theta_{Y_1}^3 + \theta_{Y_1}^5 + \theta_{Y_1}^6 + \theta_{Y_1}^7
\end{aligned}$$

(9)

Also, $\theta_{Y_2}^1, \ldots, \theta_{Y_2}^7$ and $I_{E_1^d \to Y_2}, \ldots, I_{E_{13}^d \to Y_2}$ are given similar to $\theta_{Y_1}^1, \ldots, \theta_{Y_1}^7$ and $I_{E_1^d \to Y_1}, \ldots, I_{E_{13}^d \to Y_1}$, respectively, except $Y_1$ should be replaced by $Y_2$ and $U_i$ by $V_i$, $i = 0,1,2$, everywhere. Moreover, $\mathcal{P}_i^{IN-GMS}$ denotes the set of all joint PDFs $P_{W_0 U_0 V_0 W_1 U_1 V_1 W_2 U_2 V_2 X_1 X_2}$ satisfying:

$$P_{W_0 U_0 V_0 W_1 U_1 V_1 W_2 U_2 V_2 X_1 X_2} = P_{W_0 U_0 V_0} P_{X_1 W_1 U_1 V_1|W_0 U_0 V_0} P_{X_2 W_2 U_2 V_2|W_0 U_0 V_0}$$

(10)

In Fig. 6, every two codewords connected by a directed edge are arranged in a superposition manner: The codeword at the beginning of the edge is the cloud center and the codeword at the end of the edge is the satellite. For example, the codeword $W_0^n$ is the cloud center for all other codewords. Also, in addition to $W_0^n$, both the codewords $W_1^n$ and $U_0^n$ are cloud centers for the codeword $U_1^n$. In other words, the codeword $U_1^n$ is superimposed on three codewords $W_0^n, W_1^n, U_0^n$, where $W_0^n$ itself is also a cloud center for both $W_1^n, U_0^n$. Other relations among generated codewords can be understood from Fig. 6, similarly.

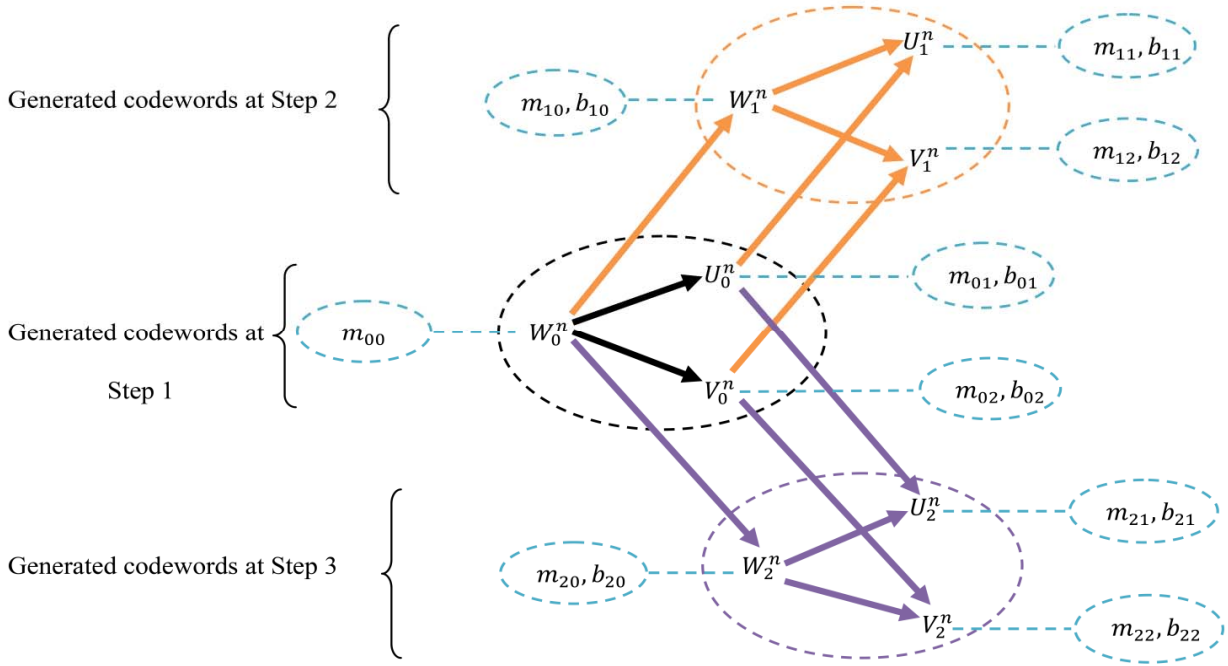

Figure 6. The graphical illustration of the generated codewords for the IN-GMS. This figure depicts the superposition structures among the generated codewords. The ellipses beside each codeword show what contains that codeword, in addition to those ones in its cloud centers.

Figure 6 clearly depicts the systematic combination of the capacity achieving scheme for the MAC with common message with the Marton's coding for the BC with common message. The superposition structures among the generated codewords is such that each $(WUV)$-triple (which performs a broadcasting) is configured in the Marton's scheme, while the triplets $(W_0 U_0 V_0)$, $(W_1 U_1 V_1)$, $(W_2 U_2 V_2)$ are configured in the capacity achieving scheme for the MAC with common message. This representation is also useful to clarify the factorization of the joint PDFs in consideration of which the resulting rate region is evaluated (10). This factorization is derived as follows:

*Each RV in the capacity achieving scheme for the MAC with common messages is replaced by the situated broadcasts RVs.*

Using this general direction, the factorization (10) can be perceived by the joint PDFs respective to the regions (1) and (2).

In the following, we describe the random codebook generation in Steps 2 and 3, in details.

*Step 2:* In this step the messages $\{M_{10}, M_{11}, M_{12}\}$ which are sent by transmitter 1, are encoded. Fix the PDFs $P_{W_1|W_0}, P_{U_1|W_1 U_0 W_0}$, and $P_{V_1|W_1 V_0 W_0}$. Let $(B_{10}, B_{11}, B_{12}) \in \mathbb{R}_+^3$ be a nonnegative triple of real numbers.

1. For each $W_0^n(m_{00})$, where $m_{00} \in [1:2^{nR_{00}}]$, generate at random $2^{n(R_{10}+B_{10})}$ independent codewords $W_1^n$ according to $Pr(w_1^n) = \prod_{t=1}^n P_{W_1|W_0}(w_{1,t}|w_{0,t}(m_{00}))$. Label these codewords $W_1^n(m_{00}, m_{10}, b_{10})$, where $m_{10} \in [1:2^{nR_{10}}]$ and $b_{10} \in [1:2^{B_{10}}]$.

2. For each triple codewords $(W_0^n(m_{00}), U_0^n(m_{00}, m_{01}, b_{01}), W_1^n(m_{00}, m_{10}, b_{10}))$, where $m_{00} \in [1:2^{nR_{00}}], m_{01} \in [1:2^{nR_{01}}], b_{01} \in [1:2^{nB_{01}}], m_{10} \in [1:2^{nR_{10}}]$, and $b_{10} \in [1:2^{nB_{10}}]$, randomly generate $2^{n(R_{11}+B_{11})}$ independent codewords $U_1^n$ according to:

$$Pr(u_1^n) = \prod_{t=1}^n P_{U_1|W_1 U_0 W_0}(u_{1,t}|w_{1,t}, u_{0,t}, w_{0,t})$$

Label these codewords $U_1^n(m_{00}, m_{01}, b_{01}, m_{10}, b_{10}, m_{11}, b_{11})$, where $m_{11} \in [1:2^{nR_{11}}]$ and $b_{11} \in [1:2^{nB_{11}}]$.

3. For each triple codewords $(W_0^n(m_{00}), V_0^n(m_{00}, m_{02}, b_{02}), W_1^n(m_{00}, m_{10}, b_{10}))$, where $m_{00} \in [1:2^{nR_{00}}], m_{02} \in [1:2^{nR_{02}}], b_{02} \in [1:2^{nB_{02}}], m_{10} \in$ $[1:2^{nR_{10}}]$, and $b_{10} \in [1:2^{nB_{10}}]$, randomly generate $2^{n(R_{12}+B_{12})}$ independent codewords $V_1^n$ according to:

$$Pr(v_1^n) = \prod_{t=1}^n P_{V_1|W_1 V_0 W_0}(v_{1,t}|w_{1,t}, v_{0,t}, w_{0,t})$$

Label these codewords $V_1^n(m_{00}, m_{02}, b_{02}, m_{10}, b_{10}, m_{12}, b_{12})$, where $m_{12} \in [1:2^{nR_{12}}]$ and $b_{12} \in [1:2^{nB_{12}}]$.

Given $(m_{10}, m_{11}, m_{12})$, define the triple $(b_{10}^T, b_{11}^T, b_{12}^T)$ as follows:

$$(b_{10}^T, b_{11}^T, b_{12}^T) \triangleq$$

$$\min \Lambda_3 \left\{ \begin{array}{c} (b_{10}, b_{11}, b_{12}), b_{1i} \in [1:2^{nB_{0i}}], i = 0,1,2: \\ \begin{pmatrix} W_0^n(m_{00}), \\ U_0^n(m_{00}, m_{01}, b_{01}^T), V_0^n(m_{00}, m_{02}, b_{02}^T), \\ W_1^n(m_{00}, m_{10}, b_{10}), \\ U_1^n(m_{00}, m_{01}, b_{01}^T, m_{10}, b_{10}, m_{11}, b_{11}), \\ V_1^n(m_{00}, m_{02}, b_{02}^T, m_{10}, b_{10}, m_{12}, b_{12}) \end{pmatrix} \in \mathcal{T}_\epsilon^n \right\}$$

(11)

In other words, $(b_{10}^T, b_{11}^T, b_{12}^T)$ is the minimum triple $(b_{10}, b_{11}, b_{12})$ (with respect to $\Lambda_3$) such that the codewords $U_0^n, V_0^n, W_0^n, W_1^n, U_1^n, V_1^n$ are jointly typical with respect to the PDF $P_{W_0 U_0 V_0 W_1 U_1 V_1}$. If there is no such triple codewords, then $(b_{10}^T, b_{11}^T, b_{12}^T) \triangleq (1,1,1)$. Note that, in the definition (11), the pair $(b_{01}^T, b_{02}^T)$ have been defined in Step 1 by (5).

In this step, the designated codewords are as:

$$\begin{pmatrix} W_1^n(m_{00}, m_{10}, b_{10}^T), \\ U_1^n(m_{00}, m_{01}, b_{01}^T, m_{10}, b_{10}^T, m_{11}, b_{11}^T), \\ V_1^n(m_{00}, m_{02}, b_{02}^T, m_{10}, b_{10}^T, m_{12}, b_{12}^T) \end{pmatrix}$$

(12)

Using the lemma proved in [3, p 15-40], we can select the sizes of the bins $B_{10}, B_{11}, B_{12}$ sufficiently large to guarantee that the codeowrds (12) are jointly typical with those in (6), with respect to the PDF $P_{W_0 U_0 V_0 W_1 U_1 V_1}$.

Given the messages $(m_{00}, m_{01}, m_{02}, m_{10}, m_{11}, m_{12})$, the first transmitter generates a codewords $X_1^n$ superimposing on the codewords (6) and (12), according to:

$$Pr(x_1^n) = \prod_{t=1}^{n} P_{X_1|W_1U_1V_1W_0U_0V_0}(x_{1,t}|w_{1,t}, u_{1,t}, v_{1,t}, w_{0,t}, u_{0,t}, v_{0,t})$$

and sends it over the channel.

*Step 3:* In this step the messages $\{M_{20}, M_{21}, M_{22}\}$ which are sent by transmitter 2, are encoded. The encoding scheme is exactly similar to Step 2. Here, the random codewords are generated based on the PDFs $P_{W_2|W_0}, P_{U_2|W_2U_0W_0}, P_{V_2|W_2V_0W_0}$, and by following the same lines as in Step 2 where the messages $m_{10}, m_{11}, m_{12}$ are replaced by $m_{20}, m_{21}, m_{22}$, the RVs $W_1, U_1, V_1, X_1$ by $W_2, U_2, V_2, X_2$, the triple $(B_{10}, B_{11}, B_{12}) \in \mathbb{R}_+^3$ by $(B_{20}, B_{21}, B_{22}) \in \mathbb{R}_+^3$, and the indices $b_{10}, b_{11}, b_{12}$ by $b_{20}, b_{21}, b_{22}$, respectively. We omit the details to avoid repetition.

*Decoding steps:* Each decoder uses a jointly typical decoder to decode its respective codewords. The encoding procedure at each receiver is as follows:

1. At receiver 1, assume that the sequence $Y_1^n$ has been received. The decoder tries to find a unique 11-tuple $(\hat{m}_{00}, \hat{m}_{01}, \hat{b}_{01}, \hat{m}_{10}, \hat{b}_{10}, \hat{m}_{11}, \hat{b}_{11}, \hat{m}_{20}, \hat{b}_{20}, \hat{m}_{21}, \hat{b}_{21})$ such that:

$$\begin{pmatrix} W_0^n(\hat{m}_{00}), U_0^n(\hat{m}_{00}, \hat{m}_{01}, \hat{b}_{01}), \\ W_1^n(\hat{m}_{00}, \hat{m}_{10}, \hat{b}_{10}), U_1^n(\hat{m}_{00}, \hat{m}_{01}, \hat{b}_{01}, \hat{m}_{10}, \hat{b}_{10}, \hat{m}_{11}, \hat{b}_{11}), \\ W_2^n(\hat{m}_{00}, \hat{m}_{20}, \hat{b}_{20}), U_2^n(\hat{m}_{00}, \hat{m}_{01}, \hat{b}_{01}, \hat{m}_{20}, \hat{b}_{20}, \hat{m}_{21}, \hat{b}_{21}), \\ Y_1^n \end{pmatrix}$$
$$\in \mathcal{T}_\epsilon^n(P_{W_0U_0W_1U_1W_2U_2Y_1})$$
(13)

If there exists such 11-tuple, then the decoder estimates its respective transmitted messages by the corresponding $(\hat{m}_{00}, \hat{m}_{01}, \hat{m}_{10}, \hat{m}_{11}, \hat{m}_{20}, \hat{m}_{21})$. If there is no such 11-tuple or there is more than one, then the decoder produces an arbitrary output and declares an error.

2. Similarly, at receiver 2 assume that the sequence $Y_2^n$ has been received. The decoder tries to find a unique 11-tuple $(\hat{m}_{00}, \hat{m}_{02}, \hat{b}_{02}, \hat{m}_{10}, \hat{b}_{10}, \hat{m}_{12}, \hat{b}_{12}, \hat{m}_{20}, \hat{b}_{20}, \hat{m}_{22}, \hat{b}_{22})$ such that:

$$\begin{pmatrix} W_0^n(\hat{m}_{00}), V_0^n(\hat{m}_{00}, \hat{m}_{02}, \hat{b}_{02}), \\ W_1^n(\hat{m}_{00}, \hat{m}_{10}, \hat{b}_{10}), V_1^n(\hat{m}_{00}, \hat{m}_{02}, \hat{b}_{02}, \hat{m}_{10}, \hat{b}_{10}, \hat{m}_{12}, \hat{b}_{12}), \\ W_2^n(\hat{m}_{00}, \hat{m}_{20}, \hat{b}_{20}), V_2^n(\hat{m}_{00}, \hat{m}_{02}, \hat{b}_{02}, \hat{m}_{20}, \hat{b}_{20}, \hat{m}_{22}, \hat{b}_{22}), \\ Y_2^n \end{pmatrix}$$
$$\in \mathcal{T}_\epsilon^n(P_{W_0V_0W_1V_1W_2V_2Y_2})$$
(14)

If there exists such 11-tuple, then the decoder estimates its respective transmitted messages by the corresponding $(\hat{m}_{00}, \hat{m}_{01}, \hat{m}_{10}, \hat{m}_{11}, \hat{m}_{20}, \hat{m}_{21})$. If there is no such 11-tuple or there is more than one, then the decoder produces an arbitrary output and declares an error.

*Analysis of error probability:* Let $0 < \epsilon < p_{min}(P_{W_0U_0V_0W_1U_1V_1W_2U_2V_2X_1X_2Y_1Y_2})$. Denote $P_{E \to Y_i}^n$ as the average error probability of decoding at the $i^{th}$ receiver, $i = 1,2$. Also, denote $P_E^n$ as the total average probability of the code. Therefore, we have:

$$P_E^n \leq P_{E \to Y_1}^n + P_{E \to Y_2}^n$$
(15)

Due to symmetry of the problem it is only required to analyze the error probability at the first receiver. The necessary conditions for vanishing the average probability of error at the first receiver can be readily extended to the second receiver by exchanging some of the parameters, as stated in the characterization of the rate region $\mathfrak{R}_i^{IN-GMS}$ given by (7). The details of the analysis of error probability in decoding at the first receiver can be found in [11]. ∎

Next, we present a class of INs-GMS for which the achievable rate region derived in Theorem 1 is optimal which yields the capacity.

*Definition:* The IN-GMS is said to be orthogonal if the alphabets transmitters are of the form $\mathcal{X}_i = \mathcal{X}_{A_i} \times \mathcal{X}_{B_i}, i = 1,2$, and the channel transition probability function satisfies:

$$\mathbb{P}(y_1, y_2|x_1, x_2) = \mathbb{P}(y_1|x_{A_1}, x_{A_2})\mathbb{P}(y_2|x_{B_1}, x_{B_2})$$
(16)

In the following theorem, we provide a full characterization of the capacity region of the orthogonal IN-GMS.

***Theorem 2)*** The capacity region of the orthogonal IN-GMS (16) denoted by $\mathfrak{C}_{IN-GMS}^{orth}$, is given as follows:

$$\mathfrak{C}_{IN-GMS}^{orth} = \bigcup_{\mathcal{P}_{IN-GMS}^{orth}} \begin{Bmatrix} (R_{00}, R_{01}, R_{02}, R_{10}, R_{11}, R_{12}, R_{20}, R_{21}, R_{22}) \in \mathbb{R}_+^9: \\ R_{10} + R_{11} \leq I(X_{A_1}; Y_1|X_{A_2}, W) \\ R_{20} + R_{21} \leq I(X_{A_2}; Y_1|X_{A_1}, W) \\ R_{10} + R_{11} + R_{20} + R_{21} \leq I(X_{A_1}, X_{A_2}; Y_1|W) \\ R_{00} + R_{01} + R_{10} + R_{11} + R_{20} + R_{21} \leq I(X_{A_1}, X_{A_2}; Y_1) \\ R_{10} + R_{12} \leq I(X_{B_1}; Y_2|X_{B_2}, W) \\ R_{20} + R_{22} \leq I(X_{B_2}; Y_2|X_{B_1}, W) \\ R_{10} + R_{12} + R_{20} + R_{22} \leq I(X_{B_1}, X_{B_2}; Y_2|W) \\ R_{00} + R_{02} + R_{10} + R_{12} + R_{20} + R_{22} \leq I(X_{B_1}, X_{B_2}; Y_2) \end{Bmatrix}$$
(17)

where $\mathcal{P}_{IN-GMS}^{orth}$ denotes the set of all joint PDFs as:

$$P_W P_{X_{A_1}|W} P_{X_{A_2}|W} P_{X_{B_1}|W} P_{X_{B_2}|W}$$
(18)

*Remarks:*

1. The rate region $\mathfrak{C}_{IN-GMS}^{orth}$ given by (17), is convex.

2. Theorem 2 shows that the essential foundation of the orthogonal IN-GMS is combined of two MACs with common message, and the capacity achieving scheme for this channel is based on a twin treatment of the superposition coding applied in [4] for the MAC with common information. This also evidences that the MAC with common message is one of the main building blocks of IN-GMS.

*Proof of Theorem 2)* To prove the direct part, we make use of the achievable rate region for the IN-GMS given in (7). By setting:

$$W_0 \equiv W_1 \equiv W_2 \equiv \emptyset, \quad U_1 \equiv X_{A_1},$$
$$V_1 \equiv X_{B_1}, \quad U_2 \equiv X_{A_2}, \quad V_2 \equiv X_{B_2}$$

in (7) and restricting the joint PDF (10) as follows:

$$P_{U_0V_0X_{A_1}X_{B_1}X_{A_2}X_{B_2}} = P_{U_0}P_{V_0}P_{X_{A_1}|U_0}P_{X_{B_1}|V_0}P_{X_{A_2}|U_0}P_{X_{B_2}|V_0}$$
(19)

we derive the following achievable rate region for the channel:

$$\bigcup_{\mathcal{P}} \begin{Bmatrix} (R_{00}, R_{01}, R_{02}, R_{10}, R_{11}, R_{12}, R_{20}, R_{21}, R_{22}) \in \mathbb{R}_+^9: \\ R_{10} + R_{11} \leq I(X_{A_1}; Y_1|X_{A_2}, U_0) \\ R_{20} + R_{21} \leq I(X_{A_2}; Y_1|X_{A_1}, U_0) \\ R_{10} + R_{11} + R_{20} + R_{21} \leq I(X_{A_1}, X_{A_2}; Y_1|U_0) \\ R_{00} + R_{01} + R_{10} + R_{11} + R_{20} + R_{21} \leq I(U_0, X_{A_1}, X_{A_2}; Y_1) \\ R_{10} + R_{12} \leq I(X_{B_1}; Y_2|X_{B_2}, V_0) \\ R_{20} + R_{22} \leq I(X_{B_2}; Y_2|X_{B_1}, V_0) \\ R_{10} + R_{12} + R_{20} + R_{22} \leq I(X_{B_1}, X_{B_2}; Y_2|V_0) \\ R_{00} + R_{02} + R_{10} + R_{12} + R_{20} + R_{22} \leq I(V_0, X_{B_1}, X_{B_2}; Y_2) \end{Bmatrix}$$
(20)

where $\mathcal{P}$ denotes the set of all joint PDFs as in (19).

Now, consider the rate region (17). Given the PDFs $P_W, P_{X_{A_1}|W}, P_{X_{A_2}|W}, P_{X_{B_1}|W}, P_{X_{B_2}|W}$, with $W \in \mathcal{W}$, define two independent RVs $U_0$ and $V_0$ with the range of $\mathcal{W}$ as follows:

$\forall a \in \mathcal{W}, i = 1,2$:
$$P_{U_0}(a) \triangleq P_{V_0}(a) \triangleq P_W(a)$$
$$P_{X_{A_i}|U_0}(.|a) \equiv P_{X_{A_i}|W}(.|a)$$
$$P_{X_{B_i}|V_0}(.|a) \equiv P_{X_{B_i}|W}(.|a)$$
(21)

Then, by substituting $P_{U_0}, P_{V_0}, P_{X_{A_1}|U_0}, P_{X_{A_2}|U_0}, P_{X_{B_1}|V_0}, P_{X_{B_2}|V_0}$ as defined by (21) in (20), the resulting rate region is equivalent to (17). The converse part will be given in [9]. ∎

Now, by an example we show that how one can establish capacity inner bounds for different sub-channels of IN-GMS using the general achievability scheme presented in Theorem 1 for this network, as well as the ***rate splitting*** technique.

Consider the two-user CIC in Fig. 7 as one of the sub-channels of the IN-GMS. We aim at extracting the HK achievable rate region [2] for this channel from the coding scheme presented for the IN-GMS. To this end, as depicted in Fig. 7, each of the messages $M_1$ and $M_2$ and thereby their respective communication rates $R_1$ and $R_2$ are split in two parts:

$$M_i = (M_{i0}, M_{ii}), \qquad R_i = R_{i0} + R_{ii}, \qquad i = 1,2$$
(22)

Now, consider the achievability scheme presented in Theorem 1 for the IN-GMS. In this coding scheme, let us restrict our attention to the case of communicating only the messages $M_{10}, M_{11}$ at transmitter 1 and the messages $M_{20}, M_{22}$ at transmitter 2. Therefore, in the rate region (7) the communication rates respective to the other messages, i.e., $M_{00}, M_{01}, M_{02}, M_{12}, M_{21}$, as well as the auxiliary RVs used to encode them (except $W_0$) are nullified. The RV $W_0$ is used to serve as the time-sharing parameter. Accordingly, in the rate region (7) we set:

$$R_{00} = R_{01} = R_{02} = R_{12} = R_{21} = 0$$
$$U_0 \equiv V_0 \equiv V_1 \equiv U_2 \equiv \emptyset$$
$$W_0 \equiv Q$$
(23)

Thereby, the distribution of the remaining RVs is given by:

$$P_{QW_0W_1U_1W_2V_2X_1X_2} = P_Q P_{X_1W_1U_1|Q} P_{X_2W_2V_2|Q}$$
(24)

On the one hand, by this assumptions we can set all the binning rates, i.e., $B_{01}, B_{02}, B_{10}, B_{11}, B_{12}, B_{20}, B_{21}, B_{22}$, equal to zero. One can easily verify that the resulting rate region by these conditions is described by the following constraints:

$$R_{11} \leq I(U_1; Y_1|W_1, W_2, Q)$$
$$R_{10} + R_{11} \leq I(W_1, U_1; Y_1|W_2, Q)$$
$$R_{20} \leq I(W_2; Y_1|W_1, U_1, Q)$$
$$R_{11} + R_{20} \leq I(U_1, W_2; Y_1|W_1, Q)$$
$$R_{10} + R_{11} + R_{20} \leq I(W_1, U_1, W_2; Y_1|Q)$$

$$R_{22} \leq I(V_2; Y_2|W_1, W_2, Q)$$
$$R_{20} + R_{22} \leq I(W_2, V_2; Y_2|W_1, Q)$$
$$R_{10} \leq I(W_1; Y_2|W_2, V_2, Q)$$
$$R_{10} + R_{22} \leq I(W_1, V_2; Y_2|W_2, Q)$$
$$R_{10} + R_{20} + R_{22} \leq I(W_1, W_2, V_2; Y_1|Q)$$
(25)

The union of all rates $(R_{10}, R_{11}, R_{20}, R_{22}) \in \mathbb{R}_+^4$ satisfying (25), taken over the set of joint PDFs as (24), is achievable for the two-user interference channel in which $M_{ii}$ is correctly decoded at the $i^{th}$ receiver and $M_{i0}$ at both receivers, $i = 1,2$. Then, note that for the two-user CIC, according to (22), $M_{10}$ is a part of the first transmitter message and $M_{20}$ a part of the second transmitter message, and hence it is required to decode only at their respective receiver, correctly. On the one hand, the constraints $R_{20} \leq I(W_2; Y_1|W_1, U_1, Q)$ and $R_{10} \leq I(W_1; Y_2|W_2, V_2, Q)$ given in (25) are the cost we have to paid to correctly decode $M_{20}$ at receiver 1 and $M_{10}$ at receiver 2, respectively. Hence, one can remove these constraints from (25) and take the others with definitions (22) as an achievable rate region for the two-user CIC.

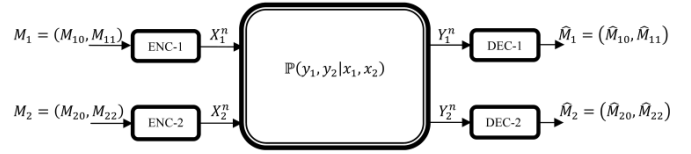

Figure 7. The two-user Classical Interference Channel (CIC).

Then, by setting $U_1 = X_1$ and $V_2 = X_2$ in the resulting rate region and applying Fourier-Motzkin elimination to remove $R_{10}, R_{11}, R_{20}, R_{22}$, the HK achievable rate region for the two-user CIC is derived, (see also [14]).

Note that the procedure described above to derive the HK rate region for the two-user CIC can also be followed for other sub-channels of IN-GMS. We follow this approach for other channel models in [9] and derive capacity inner bounds for new communication scenarios.

## CONCLUSION

In this paper, we introduced the IN-GMS and proposed an achievability scheme for it using the random coding. This scheme is systematically built based on the capacity achieving scheme for the MAC with common message as well as the best known achievability scheme for the BC with common message. We also provided a graphical illustration of the random codebook construction procedure, by using which the achievability scheme is easily understood. Moreover, we proved that the resulting rate region is optimal for a class of orthogonal INs-GMS, which yields the capacity region. Finally, we demonstrated that how this general achievability scheme can be used to derive capacity inner bounds for interference networks with different distribution of messages. In ongoing work [9], we investigate our achievable rate region for interference networks with different distribution of messages.

# APPENDIX

> *Analysis of error probability for the proposed coding scheme in Theorem I*

Given the 9-tuple $(m_{00}, m_{01}, m_{02}, m_{10}, m_{11}, m_{12}, m_{20}, m_{21}, m_{22})$ where $m_{ij} \in [1: 2^{nR_{ij}}], i,j = 0,1,2$, the encoding error events $E_1^e, E_2^e, E_3^e, E_4^e$, and also the decoding error events at the first receiver $E_0^d, E_1^d, \ldots, E_{13}^d$, are defined as follows:

*Encoding errors:*

$$E_1^e \triangleq \left\{ \begin{array}{l} \forall (b_{01}, b_{02}) \in [1: 2^{nB_{01}}] \times [1: 2^{nB_{02}}]: \\ \begin{pmatrix} W_0^n(m_{00}), \\ U_0^n(m_{00}, m_{01}, b_{01}), V_0^n(m_{00}, m_{02}, b_{02}) \end{pmatrix} \notin \mathcal{T}_\epsilon^n(P_{W_0 U_0 V_0}) \end{array} \right\}$$

(26)

$$E_2^e \triangleq \left\{ \begin{array}{l} \forall (b_{10}, b_{11}, b_{12}) \in [1: 2^{nB_{10}}] \times [1: 2^{nB_{11}}] \times [1: 2^{nB_{12}}]: \\ \begin{pmatrix} W_0^n(m_{00}), \\ U_0^n(m_{00}, m_{01}, b_{01}^\mathcal{T}), V_0^n(m_{00}, m_{02}, b_{02}^\mathcal{T}), \\ W_1^n(m_{00}, m_{10}, b_{10}), \\ U_1^n(m_{00}, m_{01}, b_{01}^\mathcal{T}, m_{10}, b_{10}, m_{11}, b_{11}), \\ V_1^n(m_{00}, m_{02}, b_{02}^\mathcal{T}, m_{10}, b_{10}, m_{12}, b_{12}) \end{pmatrix} \notin \mathcal{T}_\epsilon^n(P_{W_0 U_0 V_0 W_1 U_1 V_1}) \end{array} \right\}$$

(27)

$$E_3^e \triangleq \left\{ \begin{array}{l} \forall (b_{20}, b_{21}, b_{22}) \in [1: 2^{nB_{20}}] \times [1: 2^{nB_{21}}] \times [1: 2^{nB_{22}}]: \\ \begin{pmatrix} W_0^n(m_{00}), \\ U_0^n(m_{00}, m_{01}, b_{01}^\mathcal{T}), V_0^n(m_{00}, m_{02}, b_{02}^\mathcal{T}), \\ W_2^n(m_{00}, m_{20}, b_{20}), \\ U_2^n(m_{00}, m_{01}, b_{01}^\mathcal{T}, m_{20}, b_{20}, m_{21}, b_{21}), \\ V_2^n(m_{00}, m_{02}, b_{02}^\mathcal{T}, m_{20}, b_{20}, m_{22}, b_{22}) \end{pmatrix} \notin \mathcal{T}_\epsilon^n(P_{W_0 U_0 V_0 W_2 U_2 V_2}) \end{array} \right\}$$

(28)

$$E_4^e \triangleq \left\{ \begin{pmatrix} W_0^n(m_{00}), \\ U_0^n(m_{00}, m_{01}, b_{01}^\mathcal{T}), V_0^n(m_{00}, m_{02}, b_{01}^\mathcal{T}), \\ W_1^n(m_{00}, m_{10}, b_{10}^\mathcal{T}), \\ U_1^n(m_{00}, m_{01}, b_{01}^\mathcal{T}, m_{10}, b_{10}^\mathcal{T}, m_{11}, b_{11}^\mathcal{T}), \\ V_1^n(m_{00}, m_{02}, b_{02}^\mathcal{T}, m_{10}, b_{10}^\mathcal{T}, m_{12}, b_{12}^\mathcal{T}), \\ W_2^n(m_{00}, m_{20}, b_{20}^\mathcal{T}), \\ U_2^n(m_{00}, m_{01}, b_{01}^\mathcal{T}, m_{20}, b_{20}^\mathcal{T}, m_{21}, b_{21}^\mathcal{T}), \\ V_2^n(m_{00}, m_{02}, b_{02}^\mathcal{T}, m_{20}, b_{20}^\mathcal{T}, m_{22}, b_{22}^\mathcal{T}), \\ X_1^n(m_{00}, m_{01}, b_{01}^\mathcal{T}, m_{10}, b_{10}^\mathcal{T}, m_{11}, b_{11}^\mathcal{T}, m_{12}, b_{12}^\mathcal{T}), \\ X_2^n(m_{00}, m_{01}, b_{01}^\mathcal{T}, m_{10}, b_{10}^\mathcal{T}, m_{11}, b_{11}^\mathcal{T}, m_{12}, b_{12}^\mathcal{T}) \end{pmatrix} \notin \mathcal{T}_\epsilon^n(P_{W_0 U_0 V_0 W_1 U_1 V_1 W_2 U_2 V_2 X_1 X_2}) \right\}$$

(29)

*Decoding errors at receiver 1:*

Two types of decoding error may be occurred at the receiver: The first one is the error event where the transmitted codewords do not satisfy the decoding condition (13). This error event, denoted by $E_0^d$, is given as follows:

$$E_0^d \triangleq \left\{ \begin{pmatrix} W_0^n(m_{00}), U_0^n(m_{00}, m_{01}, b_{01}^\mathcal{T}), \\ W_1^n(m_{00}, m_{10}, b_{10}^\mathcal{T}), U_1^n(m_{00}, m_{01}, b_{01}^\mathcal{T}, m_{10}, b_{10}^\mathcal{T}, m_{11}, b_{11}^\mathcal{T}), \\ W_2^n(m_{00}, m_{20}, b_{20}^\mathcal{T}), U_2^n(m_{00}, m_{01}, b_{01}^\mathcal{T}, m_{20}, b_{20}^\mathcal{T}, m_{21}, b_{21}^\mathcal{T}), \\ Y_1^n \end{pmatrix} \notin P_{W_0 U_0 W_1 U_1 W_2 U_2 Y_1} \right\}$$

(30)

The second type is that there exist some codewords other than the transmitted ones, which satisfy the decoding error condition (13). In other words, there exist some 11-tuple $(m_{00}^*, m_{01}^*, b_{01}^*, m_{10}^*, b_{10}^*, m_{11}^*, b_{11}^*, m_{20}^*, b_{20}^*, m_{21}^*, b_{21}^*)$ such that:

$$(m_{00}^*, m_{01}^*, b_{01}^*, m_{10}^*, b_{10}^*, m_{11}^*, b_{11}^*, m_{20}^*, b_{20}^*, m_{21}^*, b_{21}^*)$$
$$\neq$$
$$(m_{00}, m_{01}, b_{01}^\mathcal{T}, m_{10}, b_{10}^\mathcal{T}, m_{11}, b_{11}^\mathcal{T}, m_{20}, b_{20}^\mathcal{T}, m_{21}, b_{21}^\mathcal{T})$$

with

$$\begin{pmatrix} W_0^n(m_{00}^*), U_0^n(m_{00}^*, m_{01}^*, b_{01}^*), \\ W_1^n(m_{00}^*, m_{10}^*, b_{10}^*), U_1^n(m_{00}^*, m_{01}^*, b_{01}^*, m_{10}^*, b_{10}^*, m_{11}^*, b_{11}^*), \\ W_2^n(m_{00}^*, m_{20}^*, b_{20}^*), U_2^n(m_{00}^*, m_{01}^*, b_{01}^*, m_{20}^*, b_{20}^*, m_{21}^*, b_{21}^*), \\ Y_1^n \end{pmatrix} \in P_{W_0 U_0 W_1 U_1 W_2 U_2 Y_1}$$

It should be noted that when two codewords construct a superposition structure, incorrect decoding of the cloud center codeword leads to incorrect decoding of the satellite one. Consequently, using the graphical illustration in Fig. 6, one can consider 13 different decoding error events of the second type at the receiver, as described in Table 1.

| -          | $W_0^n$  | $U_0^n$          | $W_1^n$          | $U_1^n$          | $W_2^n$          | $U_2^n$          |
|------------|----------|------------------|------------------|------------------|------------------|------------------|
| -          | $m_{00}$ | $(m_{01},b_{01})$| $(m_{10},b_{10})$| $(m_{11},b_{11})$| $(m_{20},b_{20})$| $(m_{21},b_{21})$|
| $E_1^d$    | ✓        | ✓                | ✓                | *                | ✓                | ✓                |
| $E_2^d$    | ✓        | ✓                | ✓                | ✓                | ✓                | *                |
| $E_3^d$    | ✓        | ✓                | ✓                | *                | ✓                | *                |
| $E_4^d$    | ✓        | ✓                | *                | *                | ✓                | ✓                |
| $E_5^d$    | ✓        | ✓                | ✓                | ✓                | *                | *                |
| $E_6^d$    | ✓        | ✓                | *                | *                | ✓                | *                |
| $E_7^d$    | ✓        | ✓                | ✓                | *                | *                | *                |
| $E_8^d$    | ✓        | *                | ✓                | *                | ✓                | *                |
| $E_9^d$    | ✓        | ✓                | *                | *                | *                | *                |
| $E_{10}^d$ | ✓        | *                | *                | *                | ✓                | *                |
| $E_{11}^d$ | ✓        | *                | ✓                | *                | *                | *                |
| $E_{12}^d$ | ✓        | *                | *                | *                | *                | *                |
| $E_{13}^d$ | *        | *                | *                | *                | *                | *                |

Table 1. The decoding errors at receiver 1.

In this table, the sign "*" indicates incorrect decoding of the respective codeword.

*Evaluation of $P_{E \to Y_1}^n$:*

Now, for the error probability of decoding at the first receiver, i.e., $P_{E \to Y_1}^n$, we can write:

$$\begin{aligned} P_{E \to Y_1}^n &\leq \frac{1}{2^{n(\Sigma_{ij} R_{ij})}} \sum_{\substack{m_{00}, m_{01}, m_{02}, \\ m_{10}, m_{11}, m_{12}, \\ m_{20}, m_{21}, m_{22}}} Pr^*(E_1^e \cup E_2^e \cup E_3^e \cup E_4^e \cup E_0^d \cup E_1^d \cup \ldots \cup E_{13}^d) \\ &\leq \frac{1}{2^{n(\Sigma_{ij} R_{ij})}} \sum_{\substack{m_{00}, m_{01}, m_{02}, \\ m_{10}, m_{11}, m_{12}, \\ m_{20}, m_{21}, m_{22}}} \begin{pmatrix} Pr^*(E_1^e) + Pr^*(E_2^e|(E_1^e)^c) + Pr^*(E_3^e|(E_1^e)^c) \\ + Pr^*(E_4^e|(E_3^e)^c, (E_2^e)^c, (E_1^e)^c) + Pr^*(E_0^d|(E_4^e)^c) + \sum_{i=1}^{13} Pr^*(E_i^d) \end{pmatrix} \end{aligned}$$

(31)

where $Pr^*(.) \triangleq Pr^* \left( . \middle| \begin{matrix} m_{00}, m_{01}, m_{02}, \\ m_{10}, m_{11}, m_{12}, \\ m_{20}, m_{21}, m_{22} \end{matrix} \right)$, and $A^c$ denotes the complement of the set $A$. Next, we bound the summands in (31). In the following analysis, $O(\epsilon)$ denotes a deterministic function of $\epsilon$, with $O(\epsilon) \to 0$ as $\epsilon \to 0$. Also, for notational convenience, we define:

$$R_{ij}^b \triangleq R_{ij} + B_{ij}, \qquad i,j \in \{0,1,2\}, \qquad (i,j) \neq (0,0)$$

(32)

First we analysis the encoding errors. For the error event $E_1^e$, using the mutual covering lemma [3] it is readily derived $Pr^*(E_1^e) \to 0$ provided that:

$$B_{01} + B_{02} > I(U_0; V_0|W_0) + O(\epsilon)$$

(33)

Then, consider the events $E_2^e$ and $E_3^e$. To derive the conditions under which the probability of these error events vanishes, we finely exploit a multivariate covering lemma proved in [3, 15-40]. First, we restate this lemma in the following.

*Lemma 1)* [3, p. 15-40] Consider a joint PDF $P_{U_0 V_0 W_1 U_1 V_1}(u_0, v_0, w_1, u_1, v_1)$ and its marginal PDFs $P_{U_0 V_0}(u_0, v_0)$, $P_{W_1}(w_1)$, $P_{U_1|W_1 U_0}(u_1|w_1, u_0)$, and $P_{V_1|W_1 V_0}(v_1|w_1, v_0)$. Let $0 < \epsilon_1 < \epsilon_2 < p_{min}(P_{U_0 V_0 W_1 U_1 V_1})$. Also, let $(B_0, B_1, B_2) \in \mathbb{R}_+^3$ be a triple of non-negative real numbers. Given a pair of deterministic n-sequences $(u_0^n, v_0^n) \in \mathcal{T}_{\epsilon_1}^n(P_{U_0 V_0})$, a random codebook is generated as follows:

1. Randomly generate $2^{nB_0}$ independent codewords $W_1^n$ according to $Pr(w_1^n) \triangleq \prod_{t=1}^n P_{W_1}(w_{1,t})$. Label these codewords as $W_1^n(b_0)$, where $b_0 \in [1:2^{nB_0}]$.

2. For the given deterministic $n$-sequences $u_0^n$ and for each $W_1^n(b_0)$ where $b_0 \in [1:2^{nB_0}]$, randomly generate $2^{nB_1}$ independent codewords $U_1^n$ according to $Pr(u_1^n) = \prod_{t=1}^n P_{U_1|W_1 U_0}(u_{1,t}|w_{1,t}, u_{0,t})$. Label these codewords as $U_1^n(u_0^n, b_0, b_1)$ where $b_1 \in [1:2^{nB_1}]$.

3. For the given deterministic $n$-sequences $v_0^n$ and for each $W_1^n(b_0)$ where $b_0 \in [1:2^{nB_0}]$, randomly generate $2^{nB_2}$ independent codewords $V_1^n$ according to $Pr(v_1^n) = \prod_{t=1}^n P_{V_1|W_1 V_0}(v_{1,t}|w_{1,t}, v_{0,t})$. Label these codewords as $V_1^n(v_0^n, b_0, b_2)$ where $b_2 \in [1:2^{nB_2}]$.

Then, there exists $O(\epsilon) \to 0$ as $\epsilon \to 0$, where if:

$$\begin{cases} B_0 > I(U_0, V_0; W_1) + O(\epsilon) \\ B_0 + B_1 > I(U_0, V_0; W_1) + I(V_0; U_1|U_0, W_1) + O(\epsilon) \\ B_0 + B_2 > I(U_0, V_0; W_1) + I(U_0; V_1|V_0, W_1) + O(\epsilon) \\ B_0 + B_1 + B_2 > I(U_0, V_0; W_1) + I(V_0; U_1|U_0, W_1) + I(U_0, U_1; V_1|V_0, W_1) + O(\epsilon) \end{cases}$$

(34)

we have:

$$Pr\left(\bigcap_{\substack{b_i, i=0,1,2 \\ b_i \in [1:2^{nB_i}]}} \left(W_1^n(b_0), U_1^n(u_0^n, b_0, b_1), V_1^n(v_0^n, b_0, b_2)\right) \notin \mathcal{T}_{\epsilon_2}^n\left(P_{U_0 V_0 W_1 U_1 V_1}|u_0^n, v_0^n\right) \bigg| u_0^n, v_0^n\right) \xrightarrow{n\to\infty} 0$$

(35)

Note that, as a simple variation of Lemma 1, one can consider the case in which all codewords are generated superimposing on another one, e.g., $w_0^n$, (for a given joint PDF $P_{W_0 U_0 V_0 W_1 U_1 V_1}(w_0, u_0, v_0, w_1, u_1, v_1)$). In this case, for vanishing the probability (35) wherein conditioning on $(u_0^n, v_0^n)$ is now replaced by $(w_0^n, u_0^n, v_0^n)$, the mutual information functions in (34) should be reformed to contain conditioning on $W_0$. In fact, we use this variation of the lemma in proving our achievability scheme for the general IN-GMS. Considering this variation of the lemma, we have depicted the superposition structures among the generated codwords in Fig. 8.

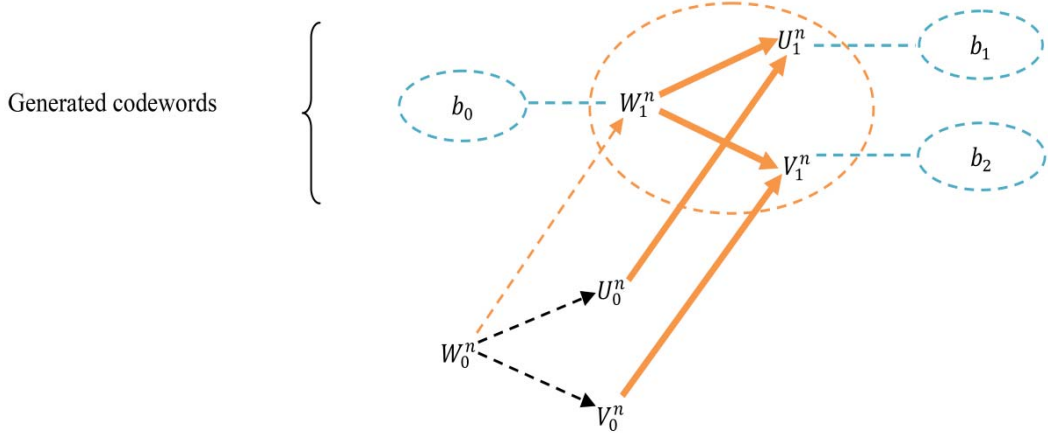

Figure 8. The graphical illustration of the generated codewords in Lemma 1. This figure depicts the superposition structures among the generated codewrods. The dashed arrows indicate the variation of the lemma where all codewords are generated superimposing on $w_0^n$.

Interestingly, the superposition structures among the codewords in Lemma 1 are exactly the same as the respective codewords of the achievability scheme in Theorem 1, as shown in Fig. 6. Therefore, we can directly apply it to evaluate $Pr^*(E_2^e|(E_1^e)^c)$ and also $Pr^*(E_3^e|(E_1^e)^c)$. We have:

$$Pr^*(E_2^e|(E_1^e)^c) = \sum_{\substack{(w_0^n, u_0^n, v_0^n) \\ \in \mathcal{T}_\epsilon^n(P_{W_0 U_0 V_0})}} Pr(w_0^n, u_0^n, v_0^n|(E_1^e)^c) \times p_{\langle w_0^n, u_0^n, v_0^n \rangle}$$

(34)

where $p_{\langle w_0^n, u_0^n, v_0^n \rangle}$ is given as follows:

$$p_{\langle w_0^n, u_0^n, v_0^n \rangle} \triangleq Pr\left(\bigcap_{\substack{b_{1i} \in [1:2^{nB_{1i}}] \\ i=0,1,2}} \left\{ \begin{pmatrix} W_1^n(m_{00}, m_{10}, b_{10}), \\ U_1^n(m_{00}, m_{01}, b_{01}^\mathcal{T}, m_{10}, b_{10}, m_{11}, b_{11}), \\ V_1^n(m_{00}, m_{02}, b_{02}^\mathcal{T}, m_{10}, b_{10}, m_{12}, b_{12}) \end{pmatrix} \notin \mathcal{T}_\epsilon^n\left(P_{W_0 U_0 V_0 W_1 U_1 V_1}\right) \right\} \bigg| w_0^n, u_0^n, v_0^n \right)$$

(35)

Using Lemma 1, one can deduce that $p_{\langle w_0^n, u_0^n, v_0^n \rangle} \to 0$ (and hence, $Pr^*(E_2^e|(E_1^e)^c) \to 0$) provided that:

$$\begin{cases} B_{10} > I(U_0, V_0; W_1|W_0) + O(\epsilon) \\ B_{10} + B_{11} > I(U_0, V_0; W_1|W_0) + I(V_0; U_1|W_0, U_0, W_1) + O(\epsilon) \\ B_{10} + B_{12} > I(U_0, V_0; W_1|W_0) + I(U_0; V_1|W_0, V_0, W_1) + O(\epsilon) \\ B_{10} + B_{11} + B_{12} > I(U_0, V_0; W_1|W_0) + I(V_0; U_1|W_0, U_0, W_1) + I(U_0, U_1; V_1|W_0, V_0, W_1) + O(\epsilon) \end{cases}$$

(36)

Symmetrically, $Pr^*(E_3^e|(E_1^e)^c) \to 0$ provided that:

$$\begin{cases} B_{20} > I(U_0, V_0; W_2|W_0) + O(\epsilon) \\ B_{20} + B_{21} > I(U_0, V_0; W_2|W_0) + I(V_0; U_2|W_0, U_0, W_2) + O(\epsilon) \\ B_{20} + B_{22} > I(U_0, V_0; W_2|W_0) + I(U_0; V_2|W_0, V_0, W_2) + O(\epsilon) \\ B_{20} + B_{21} + B_{22} > I(U_0, V_0; W_2|W_0) + I(V_0; U_2|W_0, U_0, W_2) + I(U_0, U_2; V_2|W_0, V_0, W_2) + O(\epsilon) \end{cases}$$

(37)

For the event $E_4^e$, because $X_1 W_1 U_1 V_1 \to W_0 U_0 V_0 \to X_2 W_2 U_2 V_2$ forms a Markov chain by the Markov lemma [3] we have $Pr^*(E_4^e|(E_3^e)^c, (E_2^e)^c, (E_1^e)^c) \to 0$.

Then, consider the decoding errors. For the event $E_0^d$, because $W_0 U_0 V_0 W_1 U_1 V_1 W_2 U_2 V_2 \to X_1 X_2 \to Y_1$ forms a Markov chain we have $Pr^*(E_0^d|(E_4^e)^c) \to 0$, (note that conditioning on $(E_4^e)^c$ there is no encoding error). To analyze the decoding errors indicated in Table 1, let us first evaluate the probability of the error event $E_1^d$. We have:

$$Pr^*(E_1^d) = \sum_{w_0^n, u_0^n, w_1^n, w_2^n, u_2^n} Pr^*(w_0^n, u_0^n, w_1^n, w_2^n, u_2^n, y_1^n) \times p_{\langle w_0^n, u_0^n, w_1^n, w_2^n, u_2^n, y_1^n \rangle}$$

(38)

where $p_{\langle w_0^n, u_0^n, w_1^n, w_2^n, u_2^n, y_1^n \rangle}$ is given as follows:

$p_{\langle w_0^n, u_0^n, w_1^n, w_2^n, u_2^n, y_1^n \rangle}$

$$= \sum_{\substack{(\tilde{m}_{11}, \tilde{b}_{11}) \neq \\ (m_{11}, b_{11}^T)}} Pr\left(U_1^n(m_{00}, m_{01}, b_{01}^T, m_{10}, b_{10}^T, \tilde{m}_{11}, \tilde{b}_{11}) \in \mathcal{T}_\epsilon^n(P_{W_0 U_0 W_1 U_1 W_2 U_2 Y_1}|w_0^n, u_0^n, w_1^n, w_2^n, u_2^n, y_1^n)|w_0^n, u_0^n, w_1^n\right)$$

$$= \sum_{c_{E_1^d}} \sum_{\substack{u_1^n \\ \in \mathcal{T}_\epsilon^n(P_{W_0 U_0 W_1 U_1 W_2 U_2 Y_1}|w_0^n, u_0^n, w_1^n, w_2^n, u_2^n, y_1^n)}} Pr(u_1^n|w_0^n, u_0^n, w_1^n)$$

$$\overset{(a)}{\leq} \sum_{c_{E_1^d}} 2^{nH(U_1|W_0, U_0, W_1, W_2, U_2, Y_1)(1+\epsilon)} 2^{-nH(U_1|W_0, U_0, W_1)(1-\epsilon)}$$

$$\leq 2^{n\left(R_{11}^b + H(U_1|W_0, U_0, W_1, W_2, U_2, Y_1) - H(U_1|W_0, U_0, W_1) + O(\epsilon)\right)}$$

(39)

where (a) is due to [12, Th. 1.2]. Therefore, $p_{\langle w_0^n, u_0^n, w_1^n, w_2^n, u_2^n, y_1^n \rangle} \to 0$, (and thereby $Pr^*(E_1^d) \to 0$) provided that:

$$R_{11}^b < H(U_1|W_0, U_0, W_1) - H(U_1|W_0, U_0, W_1, W_2, U_2, Y_1) - O(\epsilon) = I_{E_1^d \to Y_1} - O(\epsilon)$$

(40)

The probability of other decoding errors can be evaluated, similarly. In fact, the following general direction can be easily deduced: $Pr^*(E_i^d) \to 0$, provided that:

$$\sum_{E_i^d} \begin{pmatrix} \text{Rates respective to incorrect decoded} \\ \text{messages and bin indices} \end{pmatrix} < I_{E_i^d \to Y_1} - O(\epsilon), \quad i = 1, \dots, 13$$

(41)

where,

$I_{E_i^d \to Y_1} =$

$$\sum_{\substack{A^n \text{ is incorrctly} \\ \text{decoded}}} H\left(A \middle| \left\{ \begin{matrix} B: \\ B^n \text{ is a cloud center for } A^n \end{matrix} \right\}\right) - H\left(\left\{ \begin{matrix} C: \\ C^n \text{ is incorrectly decoded} \end{matrix} \right\} \middle| \left\{ \begin{matrix} D: \\ D^n \text{ is correctly decoded} \end{matrix} \right\}, Y_1\right)$$

(42)

Now, using the error decoding table and also the graphical illustration in Fig. 6 which depicts the superposition structures among the generated codewords, one can easily check that $I_{E_1^d \to Y_1}, \dots, I_{E_{13}^d \to Y_1}$ are given by (9). This completes the proof. ∎

It should be noted that Lemma 1 used here to analyze the encoding errors can be naturally extended to the case where the generated codewords are such that the superposition structures among them configure an arbitrary directed graph without directed cycles. This extension will be used to analyze the proposed achievability scheme for the IN-GMS with arbitrary number of transmitters and receivers [9]. Also, the general direction given by (41) and (42) to analyze the decoding errors are valid for other networks with arbitrarily large size. Using these general treatments, the derivation of the resulting achievable rate region is considerably simple [9].